\documentclass[%
 reprint,
superscriptaddress,
 amsmath,amssymb,
 aps,
 prl,
]{revtex4-2}

\usepackage{graphicx}
\usepackage{dcolumn}
\usepackage{bm}
\usepackage{hyperref}
\usepackage{cleveref}

\usepackage{physics}
\usepackage{mathrsfs}
\usepackage{amsthm}
\usepackage{mathtools}
\usepackage{xcolor}
\usepackage{ulem}
\usepackage{comment}

\newif\ifhighlighted
\highlightedfalse

\ifhighlighted
  \newcommand{\KS}[1]{\textcolor{magenta}{#1}}
  \newcommand{\KSout}[1]{\textcolor{magenta}{\sout{#1}}}
    \newenvironment{KSblock}
      {\color{magenta}}
      {}
  \newcommand{\TJK}[1]{\textcolor{blue}{#1}}
  \newcommand{\TJKout}[1]{\textcolor{blue}{\sout{#1}}}
    \newenvironment{TJKblock}
      {\color{blue}}
      {}
\else
  \newcommand{\KS}[1]{#1}
  \newcommand{\KSout}[1]{}
  
  \newcommand{\TJK}[1]{#1}
  \newcommand{\TJKout}[1]{}
  
\fi

\crefname{figure}{Fig.}{Figs.}
\crefname{equation}{Eq.}{Eqs.}

\newcommand{\GenR}{\mathcal{R}}

\newcommand{\tot}{\mathrm{tot}}
\newcommand{\hk}{\mathrm{hk}}
\newcommand{\ex}{\mathrm{ex}}
\newcommand{\iss}{\mathrm{ss}}
\newcommand{\bdg}{\mathrm{bdg}}

\begin{document}

\preprint{APS/123-QED}

\title{Optimal Driving between Nonequilibrium Steady States under a Dissipation Budget}

\author{Keisuke Sugie}
\email[E-mail me at: ]{sgekisk@sat.t.u-tokyo.ac.jp}
\affiliation{Department of Mathematical Informatics, Graduate School of Information Science and Technology, The University of Tokyo, 7-3-1 Hongo, Bunkyo-ku, Tokyo 113-8656, Japan}

\author{Dimitri Loutchko}
\affiliation{Center for Interdisciplinary AI and Data Science, Ochanomizu University, 2-1-1 Otsuka, Bunkyo-ku, Tokyo 112-8610, Japan}
\affiliation{Institute of Industrial Science, The University of Tokyo, 4-6-1, Komaba, Meguro-ku, Tokyo 153-8505 Japan}

\author{Tetsuya J. Kobayashi}%
\affiliation{Institute of Industrial Science, The University of Tokyo, 4-6-1, Komaba, Meguro-ku, Tokyo 153-8505 Japan}
\affiliation{Department of Mathematical Informatics, Graduate School of Information Science and Technology, The University of Tokyo, 7-3-1 Hongo, Bunkyo-ku, Tokyo 113-8656, Japan}

\date{\today}

\begin{abstract}

Driving between nonequilibrium steady states incurs two thermodynamic costs: excess entropy production (EP) resulting from changing control parameters, and housekeeping EP required to sustain steady currents. 
Conventional fixed-duration optimization minimizes the former while neglecting the latter. 
Here, we instead optimize excess EP over both the protocol and its duration subject to a fixed \TJK{leading order of} housekeeping-EP budget, thereby minimizing total EP for that budget \TJK{constraint}. 
In the slow-driving regime, accumulated housekeeping EP acts as an entropic time, mapping the problem to a geodesic optimization under a friction metric weighted by the steady-state EP rate. 
This resource-based geometry determines optimal paths and schedules for systems that continue to dissipate even in steady states, establishing an intrinsic trade-off between housekeeping and excess EPs. 
Near equilibrium, this framework further yields geometric lower bounds on total EP, 
\KS{attained with equipartition at the optimal budget given by a modified thermodynamic length.}
Numerical simulations of a three-state model for an ATP-driven calcium pump reveal distinct optimal protocols compared to fixed-duration driving, confirming the total-EP bounds.

\end{abstract}

\maketitle


\paragraph{Introduction.}

Optimal control in stochastic thermodynamics seeks protocols that transform mesoscopic systems at minimal thermodynamic cost  \cite{Blaber2023-ih}.
In the slow-driving regime, thermodynamic geometry provides a theoretical foundation for identifying optimal protocols to drive transitions in a fixed duration between two states both near equilibrium states (ESs) and near nonequilibrium steady states (NESSs). Unlike optimal-transport-based approaches \cite{Aurell2011-wy,Nakazato2021-at,Dechant2022-lc}, it naturally incorporates constraints on accessible parameters through parametric control, thereby yielding optimal dynamics within a physically realizable family \cite{Sivak2012-xm,Zulkowski2012-mw,Zulkowski2013-dr,Mandal2016-jd,Kamp2026-vj,Sawchuk2026-th,Zhong2022-ap,Matsuo2022-dd,Brown2020-ij,Gupta2022-as,Blaber2020-mt,Wareham2025-sk}.

This unified geometric formulation for both ESs and NESSs is achieved by adopting the Hatano--Sasa excess entropy production (EP) \cite{Hatano2001-hk,Esposito2010-id} as the thermodynamic control cost, which becomes quadratic in the protocol velocity under slow driving, yielding geodesic optimal protocols, thermodynamic-length bounds, and a Green--Kubo representation of the metric \cite{Zulkowski2013-dr,Mandal2016-jd}.

However, these earlier studies overlook the housekeeping EP required to maintain NESSs -- a limitation pointed out in \cite{Zulkowski2013-dr}.
Although minimizing the total EP does capture housekeeping dissipation, its velocity-independent contribution disrupts the purely geometric framework and can lead to qualitatively distinct optimal protocols, such as endpoint overshooting \cite{Kamp2026-vj}.
Consequently, accounting for housekeeping EP while preserving the geometry of optimal control remains an open challenge.

In addition to these technical issues, the fixed-duration formulation may not always accommodate physically appropriate constraints for transitions between NESSs.
While it is well suited for externally timed tasks such as memory erasure \cite{Schmiedl2007-pj,Proesmans2020-ou,Mulder2025-pa}, dissipation per operation may be a more natural constraint than the protocol duration for internally regulated biological systems operating far from equilibrium \cite{Cao2015-ra,Zhang2020-vq,Ouldridge2017-fh}.

Motivated by these considerations, we introduce a complementary optimal-control formulation for parametrically driven Markov jump processes, fixing the cumulative housekeeping EP rather than the protocol duration.
Specifically, we minimize the excess EP over the protocol and its duration for a fixed housekeeping EP budget:
\begin{align}
\label{eq:excess minimization with housekeeping constraint}
\min_{\tau,\bm\lambda_{0:\tau}}
\
\Sigma_\ex[\bm\lambda_{0:\tau}]
\enspace
\text{s.t.}
\enspace
\begin{aligned}
    &\Sigma_\hk[\bm\lambda_{0:\tau}] = \Sigma_\bdg,
    \\
    &
    \bm\lambda(0)=\bm\lambda_{\mathrm{i}},
    \bm\lambda(\tau)=\bm\lambda_{\mathrm{f}}
\end{aligned}
\end{align}
where the total EP, $\Sigma_\tot[\bm\lambda_{0:\tau}]=\Sigma_\bdg+\Sigma_\ex[\bm\lambda_{0:\tau}]$, is also minimized.
The accumulated housekeeping EP measures the operational progress and defines an entropic time whose terminal value is $\Sigma_\bdg$.
Under slow driving, this entropic time reduces \eqref{eq:excess minimization with housekeeping constraint} to a 
fixed-duration excess-EP minimization with a friction metric rescaled by the steady-state EPR.
This EPR rescaling also enables an exact quantification of the gap between the fixed-budget and fixed-duration excess-EP optima.
Together, the budget constraint and entropic time extend thermodynamic geometry to resource-constrained optimal control of systems dissipating continuously even at steady state.


This framework modifies conventional duration--dissipation bounds into a novel tradeoff between housekeeping and excess EPs along a trajectory, while also characterizing the gap between fixed-budget and fixed-duration excess EPs.
In the near-equilibrium regime, this framework further yields the first geometric lower bounds \eqref{eq production lower bounds} on the total EP for parametrically driven transitions between NESSs.
\KS{Budget optimization yields nonequilibrium equipartition between housekeeping and excess EP, not only for optimal driving but also within non-optimal fixed-shape protocol families.}
We illustrate differences between the formulations and confirm the total-EP bounds by numerical simulations of a three-state model for an ATP-driven calcium pump\cite{Tran2009-cc}.

\paragraph{Markov jump process and entropy productions.}
We consider a parametrically driven continuous-time Markov jump process on a finite state space $\qty{1, \cdots, N}$, where the transition rates at each time $t$ depend on a finite number of time-dependent parameters $\bm\lambda(t) = (\lambda^\mu(t))_{\mu}$.
For distinct $i\neq j$, the transition rate from state $j$ to state $i$ at $t$ is denoted by $R_{ij}(\bm\lambda(t))$, or simply by $R_{ij}$.
The diagonal elements of the transition rate matrix $\GenR = (R_{ij})$ are fixed by probability conservation, i.e. $R_{ii} = -\sum_{j \neq i} R_{ji}$.
The distribution $\bm p = (p_i)_{i=1}^N$ evolves as 
$
    \dot{\bm p}(t) = \GenR(\bm\lambda(t)) \bm p(t)
$, where the dot denotes the time derivative.
We assume that $\GenR(\bm\lambda)$ is irreducible for each $\bm\lambda$, so that the system has a unique steady distribution $\bm\pi(\bm\lambda)$ such that $\pi_i(\bm\lambda) > 0$ for any $i \in \qty{1,\cdots, N}$.
We also assume 
that every allowed transition has a corresponding reverse transition:
$
R_{ij}(\bm\lambda)>0
\Leftrightarrow
R_{ji}(\bm\lambda)>0$ for any $i\neq j$.

We introduce the probability current from $j$ to $i$,
$
    J_{ij}(t)
    =
    R_{ij}(\bm\lambda(t))p_j(t)
    -
    R_{ji}(\bm\lambda(t))p_i(t)
$.
The corresponding steady current at fixed $\bm\lambda$ is
$
    J^{\iss}_{ij}(\bm\lambda)
    =
    R_{ij}(\bm\lambda)\pi_j(\bm\lambda)
    -
    R_{ji}(\bm\lambda)\pi_i(\bm\lambda)
$.
The steady thermodynamic forces are defined as 
$
    F^{\iss}_{ij}(\bm\lambda)
    = \ln ({R_{ij}(\bm\lambda)\pi_j(\bm\lambda)} / {R_{ji}(\bm\lambda)\pi_i(\bm\lambda)})
$.
Note that $J^{\iss}_{ij}(\bm\lambda)$ and $F^{\iss}_{ij}(\bm\lambda)$ have the same sign and vanish simultaneously.
The steady state $\bm\pi(\bm\lambda)$ is an equilibrium state if $J_{ij}^\iss(\bm\lambda) = 0$ for all $i\neq j$, or equivalently, if $F_{ij}^\iss(\bm\lambda) = 0$ for all $i\neq j$.
If the steady state $\bm\pi(\bm\lambda)$  is not an equilibrium, namely if $J^{\iss}_{ij}(\bm\lambda)\neq 0$ for at least one pair of states, it is called a NESS.
{For a protocol $\bm\lambda_{0:\tau}$, we define $\delta\equiv\max_{t\in[0,\tau],\,i<j}\abs{J_{ij}^{\iss}(\bm\lambda(t))}$ and refer to $\delta=0$, $\delta\ll1$, and $\delta=\mathcal O(1)$ as equilibrium, near-equilibrium, and far-from-equilibrium driving, respectively.
}

The total entropy production rate (EPR) is given by
$
    \sigma_{\rm tot}(t)
    \equiv
    \sum_{i<j}
    J_{ij}(t)
    \ln
    ({
        R_{ij}(\bm\lambda(t))p_j(t)
    }/{
        R_{ji}(\bm\lambda(t))p_i(t)
    })
$.
Following the decomposition in steady-state thermodynamics \cite{Hatano2001-hk,Esposito2010-id}, we decompose the total EPR as
$
    \sigma_{\rm tot}(t)
    =
    \sigma_{\rm hk}(t)
    +
    \sigma_{\rm ex}(t)
$,
with the housekeeping EPR given by
$
\sigma_{\rm hk}(t)
\equiv
\sum_{i<j}J_{ij}(t)F_{ij}^{\iss}(\bm\lambda(t))
$,
and the excess EPR as the complementary part.
Note that the three EPRs are nonnegative \cite{Ge2009-cf,Esposito2010-id}.
The housekeeping EPR vanishes, $\sigma_\hk(t) = 0$, when $\bm\pi(\bm\lambda)$ is an equilibrium.
In 
the quasistatic limit $\bm p(t) \to \bm\pi(\bm\lambda(t))$, the excess EPR goes to zero: $\sigma_\ex(t) \to 0$.
The three types of entropy productions (EPs) for a protocol $\bm\lambda_{0:\tau}$ are 
$
    \Sigma_\alpha[\bm\lambda_{0:\tau}]
    \equiv
    \int_0^\tau dt\,\sigma_\alpha(t), \,
    \alpha\in\{\mathrm{tot},\mathrm{hk},\mathrm{ex}\}
$.
The total EP is also decomposed into the two contributions:
$
    \Sigma_\tot[\bm\lambda_{0:\tau}] = \Sigma_\hk[\bm\lambda_{0:\tau}] + \Sigma_\ex[\bm\lambda_{0:\tau}].
$

\begin{figure*}[t]
    \centering
    \includegraphics[width=\textwidth]{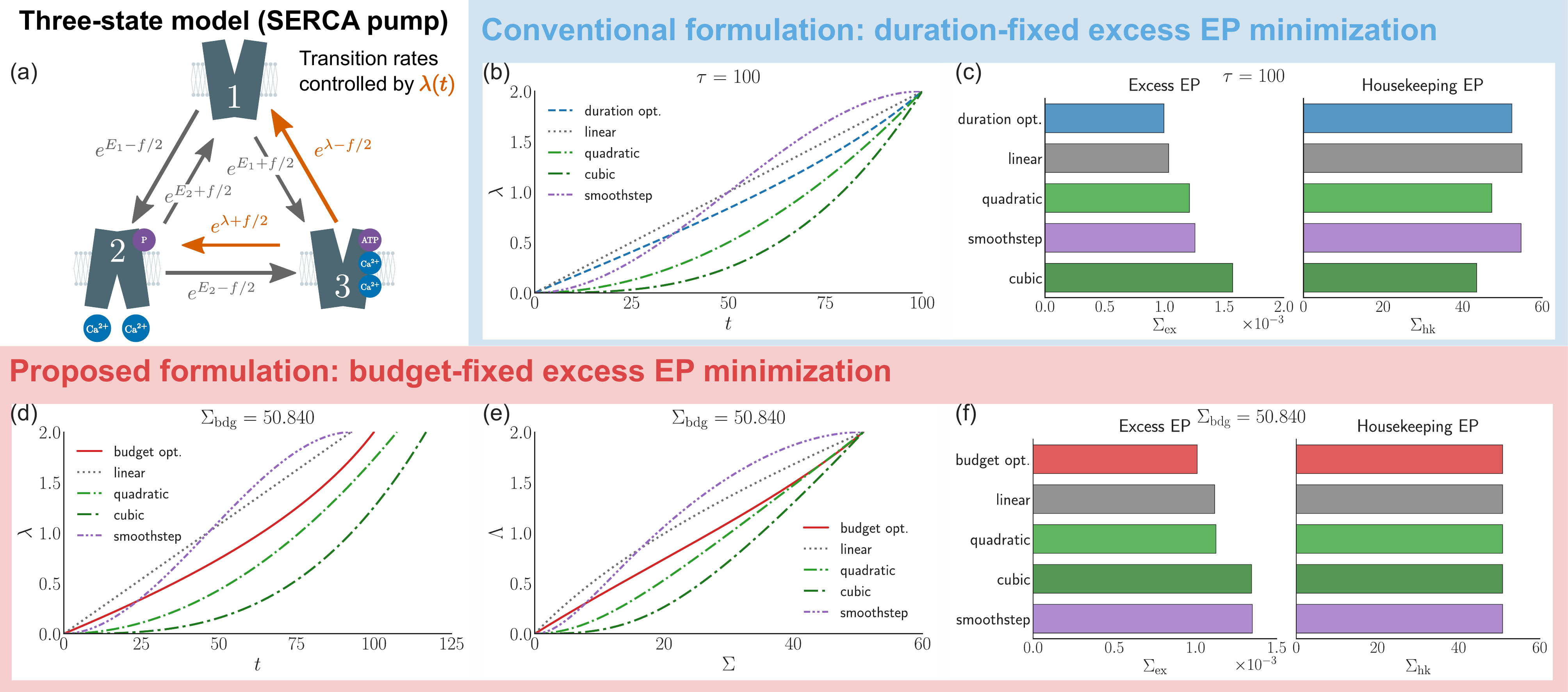}
    \caption{
    Comparison of excess-EP minimization at fixed duration and fixed housekeeping-EP budget in the three-state model. 
    (a) Three-state model as an ATP-driven calcium pump: ATP-free (1), phosphorylated/Ca$^{2+}$-releasing (2), and ATP/Ca$^{2+}$-binding (3) states; orange arrows indicate $\lambda(t)$-dependent rates representing phospholamban regulation.
    (b,c) Duration-optimal and reference protocols and their excess and housekeeping EPs at $\tau=100$. (d)--(f) Budget-optimal and reference protocols at $\Sigma_{\rm bdg}=50.840$, shown in physical time, entropic time, and EPs. Parameters: $E_1=1$, $E_2=2$, $f=0.4$, and $\lambda:0\to2$. See \cite{SM}.
    }
    \label{fig:EPs under fixed duration}
\end{figure*}

\paragraph{Slow driving regime and thermodynamic geometry.}
We now recall the finite-time slow driving approximation \cite{Mandal2016-jd,Matsuo2022-dd}. 
Suppose that the protocol duration is long but finite $\tau \gg 1$, so that the parameter velocity is small: $\|\dot{\bm\lambda}\| = \mathcal{O}(\tau^{-1})$. 
In this limit, the deviation of the distribution at time $t$ from the instantaneous steady state, 
$
    \bm p(t) - \bm\pi(\bm\lambda(t)),
$
can be approximated as
$
    \bm p(t) - \bm\pi(\bm\lambda(t)) = \GenR(\bm\lambda(t))^+  \dot{\bm \pi}(\bm\lambda(t)) + \mathcal{O}(\tau^{-2})
$.
Here, $
\GenR^+
\equiv
\left(\GenR+\bm\pi\bm 1^\top\right)^{-1}
-
\bm\pi\bm 1^\top,
$ denotes the Drazin inverse of the transition rate matrix, 
where $\bm 1$ denotes the $N$-dimensional column one vector.
Equivalent characterizations of $\GenR^+$ are provided in \cite{SM}.

In the far-from-equilibrium driving regime, substituting the slow driving approximation into the decomposed EPs gives their leading forms.
The housekeeping EP has a $\mathcal{O}(\tau)$ contribution,
\begin{align}
    \Sigma_{\rm hk}[\bm\lambda_{0:\tau}]
    =
    \int_0^\tau 
    \sigma^{\iss}(\bm\lambda(t)) \,\dd t
    +
    \mathcal{O}(1),
\end{align}
where
$
    \sigma^{\iss}(\bm\lambda)
    =
    \sum_{i<j}
    J^{\iss}_{ij}(\bm\lambda)
    F^{\iss}_{ij}(\bm\lambda)
$
is the steady state EPR at fixed $\bm\lambda$. 
The excess EP starts from first order in the driving speed and takes the quadratic form
\begin{align}
    \Sigma_{\rm ex}[\bm\lambda_{0:\tau}]
    =
    \int_0^\tau 
    \dot\lambda^\mu(t)
    \zeta_{\mu\nu}(\bm\lambda(t))
    \dot\lambda^\nu(t) \,\dd t
    +
    \mathcal{O}(\tau^{-2}),
\end{align}
where we adopt the Einstein convention for repeated Greek indices.
Here $\zeta_{\mu\nu}(\bm\lambda)$ is the thermodynamic friction matrix associated with the cross correlations in the instantaneous steady state $\bm\pi(\bm\lambda)$. 
At this order, $\Sigma_{\rm ex}\geq\mathcal L_\gamma^2/\tau$, where $\gamma$ is the protocol path in parameter space and $\mathcal L_\gamma\equiv\int_\gamma\sqrt{\dd\lambda^\mu\zeta_{\mu\nu}\dd\lambda^\nu}$ is its thermodynamic length.
See {\cite{SM}} for the definition of $\zeta$ and the derivation.

Under the slow-driving approximation, minimizing the excess EP $\Sigma_{\mathrm{ex}}[\bm\lambda_{0:\tau}]$ at fixed duration $\tau$ reduces to
\begin{align}
    \begin{aligned}
        \label{eq:approximated excess minimization under fixed duration}
        \min_{\bm\lambda_{0:\tau}}  \ \int_0^\tau \dot{\bm \lambda}^\top \zeta(\bm\lambda) \dot{\bm\lambda} \,\dd t \enspace
        \textrm{s.t.} \enspace
        \bm \lambda(0) = \bm \lambda_{\mathrm{i}},\bm \lambda(\tau) = \bm \lambda_{\mathrm{f}}.
    \end{aligned}
\end{align}
The optimal protocol, denoted by $\bm\lambda^\ast_{0:\tau}$, of this variational problem is a geodesic of the metric $\zeta_{\mu\nu}$ \cite{Zulkowski2012-mw,Sivak2012-xm,Zulkowski2013-dr,Mandal2016-jd}
and satisfies the geodesic equation $ \ddot{\lambda}^{\ast\mu} + \Gamma^\mu_{\nu\rho}(\bm\lambda^\ast) \dot{\lambda}^{\ast\nu} \dot{\lambda}^{\ast\rho} = 0$, where 
$\Gamma^\mu_{\nu\rho} = 
\zeta^{\mu\kappa} \qty( \partial_\nu \zeta_{\kappa\rho} + \partial_\rho \zeta_{\kappa\nu} - \partial_\kappa \zeta_{\nu\rho}
)/2$
denotes the Christoffel symbol associated with $\zeta_{\mu\nu}$.


\paragraph{Fixed dissipation budget formulation and entropic-time reparameterization.}
We now analyze the excess optimal driving principle under a housekeeping budget \eqref{eq:excess minimization with housekeeping constraint} in the slow finite-time and far-from-equilibrium driving regime. 
Keeping only the leading terms in the housekeeping and excess EPs, \eqref{eq:excess minimization with housekeeping constraint} is reduced to the constrained problem:
\begin{align}
    \label{eq:approximated excess minimization with housekeeping constraint}
        \min_{\tau,\bm\lambda_{0:\tau}}  \ \int_0^\tau \dot{\bm \lambda}^\top \zeta(\bm\lambda) \dot{\bm\lambda} \,\dd t \enspace
    \textrm{s.t.} \enspace 
    \begin{aligned}
        &\int_0^\tau  \sigma^\iss(\bm \lambda) \,\dd t = \Sigma_\bdg,\\
        & \bm \lambda(0) = \bm \lambda_{\mathrm{i}},\bm \lambda(\tau) = \bm \lambda_{\mathrm{f}}.
    \end{aligned}
\end{align}
This reduced formulation minimizes the leading-order excess EP under a budget on the steady-state housekeeping EP (see \cite{SM} for its physical interpretation).
The slow-driving approximation reduces the dependence of EPs on $\bm\lambda_{0:t}$ to time-local functionals of $\bm\lambda$ and $\dot{\bm\lambda}$, yielding an isoperimetric variational problem.
Although this problem can be treated using a Lagrange multiplier \cite{Van-Brunt2004-ml}, its particular constraint admits a more direct reformulation.


We use the accumulated housekeeping EP as an entropic time, thereby converting the constrained problem \eqref{eq:approximated excess minimization with housekeeping constraint} into a fixed-horizon problem.
For a given protocol $\bm\lambda_{0:\tau}$, we introduce the protocol-dependent entropic time as the leading-order housekeeping EP up to time $t$:
\begin{align}
\label{eq:entropic time}
\Sigma(t)
\equiv
\int_0^t
\sigma^\iss(\bm\lambda(s))
\,\dd s.
\end{align}
{Thus, equal increments in entropic time, $\dd \Sigma$, translate to shorter real-time intervals in regions of higher housekeeping EPR, whereas they span longer durations in regions where the EPR is lower.}
Provided that $\sigma^\iss(\bm\lambda)>0$ along the protocol, this map is invertible and sends $t \in [0, \tau]$ to $\Sigma \in [0,\Sigma_\bdg]$, where the housekeeping budget constraint becomes the fixed-horizon condition $\Sigma(\tau) = \Sigma_\bdg$.
Denoting the inverse of $\Sigma(\cdot)$ as $t(\cdot)$, we define the entropic-time protocol $\bm\Lambda_{0:\Sigma_\bdg}\equiv \qty{\bm\Lambda(\Sigma)}_{\Sigma \in[0,\Sigma_\bdg]}$ by $\bm\Lambda(\Sigma)\equiv\bm\lambda(t(\Sigma))$.
The time derivative and differential are transformed as
\begin{align}
\dot{\bm\lambda}
=
\sigma^\iss(\bm\Lambda)\bm\Lambda',
\quad
\dd t
=
\frac{\dd\Sigma}{\sigma^\iss(\bm\Lambda)},
\end{align}
where the prime $'$ denotes the differentiation with respect to ${\Sigma}$. 
With
$
\widetilde{\zeta}(\bm\Lambda)
\equiv
\sigma^\iss(\bm\Lambda) \cdot \zeta(\bm\Lambda)
$
, the leading-order excess EP is
$
\Sigma_\ex =
\int_0^\tau
\dot{\bm\lambda}^{\top}
\zeta(\bm\lambda)
\dot{\bm\lambda}
\,\dd t
=
\int_0^{\Sigma_\bdg}
{\bm\Lambda'}^{\top}
\widetilde{\zeta}(\bm\Lambda)
\bm\Lambda'
\,\dd\Sigma
$.
The problem \eqref{eq:approximated excess minimization with housekeeping constraint} is then transformed into
\begin{align}
    \label{eq:quadratic excess minimization with modified metric}
        \min_{\bm{\Lambda}_{0:\Sigma_\bdg}} \ \int_0^{\Sigma_\bdg} {\bm\Lambda'}^\top \widetilde{\zeta}(\bm\Lambda) {\bm\Lambda}' \,\dd \Sigma \enspace
        \textrm{s.t.} \enspace
    \begin{aligned}
        &\bm \Lambda(0) = \bm \lambda_{\mathrm{i}},\\
        &\bm \Lambda({\Sigma_\bdg}) = \bm \lambda_{\mathrm{f}}.
    \end{aligned}
\end{align}
Conversely, any entropic-time protocol
$\bm\Lambda_{0:\Sigma_\bdg}$
determines a corresponding physical time $t(\cdot)$ through
\begin{align}
t(\Sigma)
=
\int_0^\Sigma
\frac{\dd\xi}
{\sigma^\iss(\bm\Lambda(\xi))}.
\end{align}
The original constrained problem \eqref{eq:approximated excess minimization with housekeeping constraint} can therefore be solved by optimizing $\bm\Lambda$ over the fixed entropic-time horizon and subsequently reconstructing the real-time protocol, $\bm\lambda(t)=\bm\Lambda(\Sigma(t))$, 
with duration $\tau=t(\Sigma_\bdg)$.

Assuming that $\zeta$ is symmetric and positive definite and thus defines a metric on the control-parameter space, the reparametrization \eqref{eq:entropic time} can be interpreted as a generalized Sundman transformation \cite{Carinena2022-et} in Riemannian geometry, which induces the conformally related (i.e., differing only by multiplication by a positive function) metric $\widetilde{\zeta}$.
The transformed problem \eqref{eq:quadratic excess minimization with modified metric} then takes the same Riemannian form as the fixed-duration slow-driving problem \eqref{eq:approximated excess minimization under fixed duration}, with the modified metric $\widetilde{\zeta}$. 
The optimal protocol $\bm\Lambda^\dagger(\Sigma)$ of \eqref{eq:quadratic excess minimization with modified metric} therefore satisfies the geodesic equation in entropic time,
$
    (\Lambda^{\dagger\mu})''
    +
    \widetilde{\Gamma}^{\mu}_{\nu\rho}(\bm\Lambda^\dagger)
    (\Lambda^{\dagger\nu})'
    (\Lambda^{\dagger\rho})'
    =
    0
$,
where $\widetilde{\Gamma}^{\mu}_{\nu\rho}$ denotes the Christoffel symbols associated with $\widetilde{\zeta}$.

We numerically compare the duration- and budget-fixed formulations in the three-state model \cite{Kamp2026-vj} of \cref{fig:EPs under fixed duration}(a), using the optimal and reference protocols.
{This model also serves as a simplified description of an ATP-driven calcium pump \cite{Tran2009-cc} regulated by phospholamban \cite{Simmerman1998-bz} (see \cite{SM}).}
At fixed duration [\cref{fig:EPs under fixed duration}(b) and (c)], the optimal protocol $\bm\lambda_{0:\tau}^\ast$ has larger housekeeping and total EP than the quadratic and cubic protocols. 
At fixed budget, the protocols have different durations but terminate at the same entropic time [\cref{fig:EPs under fixed duration}(d) and (e)]. 
Unlike the total-EP-optimal protocol of Kamp and Proesmans \cite{Kamp2026-vj}, the optimal protocol $\bm\lambda^\dagger(t) \equiv \bm\Lambda^\dagger(\Sigma(t))$ of \eqref{eq:approximated excess minimization with housekeeping constraint} shows no overshoot [\cref{fig:EPs under fixed duration}(d)], because of the retained Riemannian geometry.
Since the budget fixes the housekeeping EP to leading order, $\bm\lambda^\dagger_{0:\tau}$ minimizes the excess EP [\cref{fig:EPs under fixed duration}(f)] and, in this simulation, yields the smallest total EP among the protocols considered; the EP values are listed in {\cite{SM}}.


\paragraph{Inherited geometry and its implications.}
This inherited geometry allows us to generalize conventional trade-offs and bounds between physical time and EP \cite{Crooks2007-rx,Sivak2012-xm,Mandal2016-jd}. 
For a path $\gamma$ in parameter space, the modified metric defines the thermodynamic length
\begin{align}
\widetilde{\mathcal L}_\gamma
\equiv\int_\gamma\sqrt{\dd\Lambda^\mu\widetilde{\zeta}_{\mu\nu}(\boldsymbol{\Lambda})\dd\Lambda^\nu}.
\end{align}
Applying the Cauchy--Schwarz inequality to the objective functional in \eqref{eq:quadratic excess minimization with modified metric} gives
\begin{align}
\label{eq:excess-budget-inequality}
\Sigma_{\mathrm{ex}}[\bm\lambda_{0:\tau}]\ge\frac{(\widetilde{\mathcal L}_\gamma)^2}{\Sigma_\bdg}.
\end{align}
This bound represents a trade-off between excess EP and the available housekeeping budget: for a fixed path $\gamma$, reducing the excess EP requires a larger housekeeping budget. 
The equality holds iff it is traversed at constant thermodynamic speed in entropic time, $\Lambda'^{\mu}\widetilde{\zeta}_{\mu\nu}\Lambda'^{\nu}=\mathrm{const}$; 
further optimization over $\gamma$ selects the geodesic $\gamma^\dagger$ traced by $\bm\Lambda^\dagger$.
In physical time, this condition reads $\dot{\lambda}^{\mu}\zeta_{\mu\nu}\dot{\lambda}^{\nu}/\sigma^{\iss}(\bm\lambda)=\mathrm{const}$, so the optimal parametrization maintains a constant leading-order ratio between the excess and housekeeping EPRs.

The decomposition of the usual friction metric $\zeta$ into integrated relaxation times and Fisher information \cite{Sivak2012-xm,Mandal2016-jd,Sawchuk2026-th} is updated for the modified metric:
\begin{align}
\widetilde{\zeta}_{\mu\nu}(\boldsymbol{\lambda}) 
= 
\sigma^{\iss}(\boldsymbol{\lambda}) \tau_{\mu\nu}(\boldsymbol{\lambda}) \cdot I_{\mu\nu}(\boldsymbol{\lambda}),
\label{eq:decomposition of modified metric}
\end{align}
where $\mathcal{I}=(I_{\mu\nu})$ is the Fisher information matrix of the instantaneous steady state and $\tau_{\mu\nu}$ is the symmetrized integrated relaxation time, given by
\begin{align}
I_{\mu\nu}
&\equiv C_{\mu\nu}(0) =  C_{\nu\mu}(0) =
\sum_i
\pdv{\ln \pi_i}{\lambda_\mu}
\pdv{\ln \pi_i}{\lambda_\nu}\pi_i,
\\
\tau_{\mu\nu}
&\equiv
\int_0^\infty
\frac{C_{\mu\nu}(s) + C_{\nu\mu}(s)}{C_{\mu\nu}(0) + C_{\nu\mu}(0)} \,\dd s.
\end{align}
Here, $C_{\mu\nu}(s)$ is the cross-correlation between the score functions in the instantaneous steady state,
$
C_{\mu\nu}(s)
\equiv
\sum_{i,j}
\partial_{\lambda_\mu}{\ln \pi_i}\cdot 
\partial_{\lambda_\nu}{\ln \pi_j}\cdot
(e^{s\GenR})_{ij}\pi_j
$ \cite{SM}.
This decomposition clarifies that, across the parameter-space manifold, the modified metric integrates the heterogeneity of housekeeping EPR with that of relaxation time back to the steady states.

\paragraph{Path EPR heterogeneity quantifies the excess-EP ratio.}
The impact of this heterogeneity of the steady-state EPR accumulates along the optimal path, resulting in a quantitative relationship between the optimal excess EPs for 
the optimal duration-fixed ($\bm\lambda^\ast_{0:\tau}$) and budget-fixed ($\bm\lambda^\dagger_{0:\tau}$) protocols.
We compare here the excess EPs under two optimal protocols for the same physical duration $\tau$ by exploiting the fact that there exists a unique budget whose optimum has duration $\tau$  for the budget-fixed problem \cite{SM}.
We define the excess-EP ratio as
$
\rho \equiv
{\Sigma_{\ex}[\bm\lambda^\dagger_{0:\tau}]}
/{\Sigma_{\ex}[\bm\lambda^\ast_{0:\tau}]}
\geq 1
$.
Their optimality conditions yield
\begin{align}
\rho
&=
\underbrace{
\vphantom{\qty(\frac{\mathcal L_{\gamma^\dagger}}{\mathcal L_{\gamma^\ast}})^2}
\qty{
1+
\operatorname{CV}_{\gamma^\dagger}^{2}
\qty[
\sqrt{\sigma^{\iss}}
]
}
}_{\displaystyle \rho_{\rm sch}} \cdot
\underbrace{
\vphantom{
\qty{
1+
\operatorname{CV}_{\gamma^\dagger}^{2}
\qty[
\sqrt{\sigma^{\iss}}
]
}} 
\qty(
\frac{\mathcal L_{\gamma^\dagger}}
{\mathcal L_{\gamma^\ast}}
)^2
}_{\displaystyle \rho_{\rm path}},
\label{eq}
\end{align}
where $\rho_{\rm sch}$ and $\rho_{\rm path}$ quantify time-scheduling and path contributions, with $\rho_{\rm sch}$ set by steady-state-EPR heterogeneity along the budget-optimal path and $\rho_{\rm path}$ by the thermodynamic-length ratio of $\gamma^\dagger$ to $\gamma^\ast$, the paths traced by $\bm\lambda^\dagger_{0:\tau}$ and $\bm\lambda^\ast_{0:\tau}$, respectively.
Here,
$\operatorname{CV}_{\gamma}^2[f]
\equiv
\langle(f-\langle f\rangle_\gamma)^2\rangle_\gamma/
\langle f\rangle_\gamma^2$,
with
$\langle f\rangle_\gamma\equiv\int_\gamma f\,\dd\mu_\gamma$ and
$
\dd\mu_\gamma
\equiv
Z_\gamma^{-1}
\sqrt{
{
\dd\bm\lambda^{\top}
\zeta(\bm\lambda)\dd\bm\lambda
}/{
\sigma^{\iss}(\bm\lambda)
}
}$.
For the budget-optimal protocol, $\dd\mu_{\gamma^\dagger}=\dd t/\tau$.
The rescaling of the metric by $\sigma^{\iss}(\boldsymbol{\lambda})$ encourages the budget-optimal protocol to choose a path avoiding high $\sigma^{\iss}(\boldsymbol{\lambda})$ regions and a speed scheduling to pass through such regions promptly.
The factors $\rho_{\rm path}$ and $\rho_{\rm sch}$ dissect and quantify the costs of excess EPs for the rerouting and rescheduling from $\gamma^{*}$, respectively.
The CV form of $\rho_{\rm sch}$ yields the Kantorovich-type bound \cite{Mitrinovic2012-xf} $\rho_{\rm sch}\le\cosh^2 q_{\gamma^\dagger}$, where $2q_{\gamma^\dagger}\equiv\ln(M_{\gamma^\dagger}/m_{\gamma^\dagger})$, $m_{\gamma^\dagger}\equiv\min_{\gamma^\dagger}\sqrt{\sigma^\iss}$, and $M_{\gamma^\dagger}\equiv\max_{\gamma^\dagger}\sqrt{\sigma^\iss}$.
The proof and a tighter Bhatia--Davis bound~\cite{Bhatia2000-fz} incorporating the path average $\expval{\sqrt{\sigma^\iss}}_{\gamma^\dagger}$ are given in \cite{SM}.


\begin{figure}[t]
    \centering
    \includegraphics[width=\columnwidth]{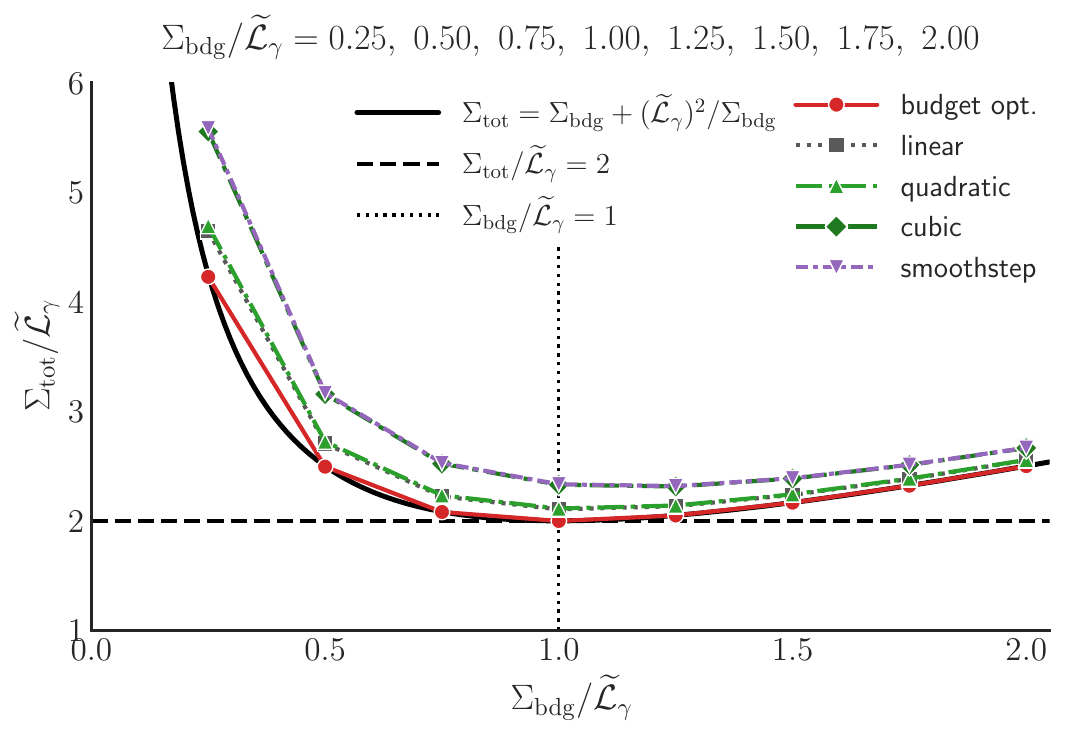}
    \caption{
    {
    Near-equilibrium verification of the total-EP bounds for the three-state model in \cref{fig:EPs under fixed duration}(a), using the protocols of \cref{fig:EPs under fixed duration}(d)--(f). 
    All protocols follow the common one-dimensional path $\gamma=[\lambda_{\rm i},\lambda_{\rm f}]=[0,2]$, and total EP and budget are normalized by $\widetilde{\mathcal L}_{\gamma}$. Solid curve: budget-dependent bound; dashed line: budget-independent bound; dotted line: optimal budget $\Sigma_{\bdg}=\widetilde{\mathcal L}_{\gamma}$. See \cite{SM} for numerical details.
    }
    }
    \label{fig:total EP bounds}
\end{figure}

\paragraph{Equilibrium and near-equilibrium driving.}
We now characterize the problem \eqref{eq:approximated excess minimization with housekeeping constraint} at and near equilibrium.

For driving equilibrium systems, i.e., $\delta = 0$, only the trivial budget $\Sigma_\bdg=0$ is feasible and the entropic-time parametrization degenerates. 
The problem \eqref{eq:approximated excess minimization with housekeeping constraint} then reduces to 
minimization without  constraint on the duration
of the total EP, whose infimum vanishes in the quasistatic limit $\tau\to\infty$, recovering classical equilibrium driving.

Near equilibrium, $\delta\ll1$, the two EPs scale as
\begin{align}
\Sigma_\hk=\mathcal O(\delta^2\tau),
\qquad
\Sigma_\ex=\mathcal O(\tau^{-1}).
\end{align}
Note that the $\mathcal{O}(\delta)$ term in $\Sigma_\hk$ vanishes identically; see {\cite{SM}}. 
Hence, the two contributions are of the same order for $\delta=\mathcal O(\tau^{-1})$, whereas housekeeping dominates for $\delta\gg\tau^{-1}$ and excess for $\delta\ll\tau^{-1}$.

At the crossover $\delta=\mathcal O(\tau^{-1})$, the two EPs are of the same order and therefore both enter the total EP. 
Previous approaches provided lower bounds only on the excess EP; our housekeeping-budget formulation instead yields a lower bound on the total EP in this crossover regime. 
Within the slow-driving approximation, the excess-EP bound \eqref{eq:excess-budget-inequality} with $\Sigma_\hk=\Sigma_\bdg$ gives, for any candidate protocol $\bm\lambda_{0:\tau}$ for problem \eqref{eq:approximated excess minimization with housekeeping constraint} with path $\gamma$,
\begin{align}
\Sigma_\tot[\bm\lambda_{0:\tau}]
&\ge
\Sigma_\bdg+\frac{(\widetilde{\mathcal L}_\gamma)^2}{\Sigma_\bdg}
\ge 2\widetilde{\mathcal L}_\gamma.
\label{eq production lower bounds}
\end{align}
The first inequality gives a budget-dependent bound on the total EP, with equality under the same condition as in \eqref{eq:excess-budget-inequality}.
The second inequality gives a budget-independent bound on the total EP, with equality iff $\Sigma_\bdg=\widetilde{\mathcal L}_\gamma$.
\KS{
Both inequalities are saturated by optimal driving along $\gamma^\ast$ at $\Sigma_\bdg=\widetilde{\mathcal L}_{\gamma^\ast}$, 
establishing a nonequilibrium equipartition theorem: the minimum total EP is shared equally between housekeeping and excess EP, $\Sigma_\hk=\Sigma_\ex=\widetilde{\mathcal L}_{\gamma^\ast}$.}

\KS{
Equipartition extends even to non-optimal fixed-shape protocol families $\bm\lambda(t)=\hat{\bm\lambda}(t/\tau)$.
Minimizing the total EP over the duration determines the family's optimal budget $\Sigma_\bdg^{\rm opt}[\hat{\bm\lambda}]$, at which $\Sigma_\hk=\Sigma_\ex$ \cite{SM}.
The equality follows because, as for optimal driving, the excess EP is inversely proportional to the budget,
$\Sigma_\ex[\bm\lambda_{0:\tau}]=(\Sigma_\bdg^{\rm opt}[\hat{\bm\lambda}])^2/\Sigma_\bdg$.
The thermodynamic length bounds this optimal budget from below,
$\Sigma_\bdg^{\rm opt}[\hat{\bm\lambda}]\geq\widetilde{\mathcal L}_\gamma$,
even for a non-geodesic path $\gamma$.
Equality holds iff the thermodynamic speed in entropic time is constant; otherwise, additional excess EP raises the optimal budget.
}

We numerically test the total-EP bounds in the near-equilibrium regime using the same three-state model and five protocols as in \cref{fig:EPs under fixed duration} [\cref{fig:total EP bounds}]. 
Consistent with our theoretical results, the total EP for every protocol lies above the budget-dependent bound $\Sigma_\bdg + (\widetilde{\mathcal L}_\gamma)^2/\Sigma_\bdg$, attained by the budget-fixed optimal protocol. 
All curves also are above the budget-independent bound $2\widetilde{\mathcal L}_\gamma$, attained by the optimal protocol under $\Sigma_{\mathrm{bdg}}=\widetilde{\mathcal L}_\gamma$. 
\KS{
For the reference families, the minimizing budgets in \cref{fig:total EP bounds} are consistent with the predicted values $\Sigma_\bdg^{\rm opt}[\hat{\bm\lambda}]$ ($1.05$--$1.15$ times $\widetilde{\mathcal L}_\gamma$) \cite{SM},
illustrating how thermodynamic geometry guides budget selection \TJK{even} for non-optimal protocols.
}

\paragraph{Summary and discussion.}
We formulate optimal driving at fixed housekeeping-EP budget and show that, under slow driving, entropic-time reparametrization yields a fixed-horizon geometric problem with metric $\sigma^{\iss}\bm\zeta$ and a budget--excess trade-off.
Integral constraints have previously been applied to dynamical activity \cite{Shiraishi2018-qd,Dechant2022-lc,Van-Vu2023-ug}, alongside edge-wise constraints \cite{Remlein2021-qt,Ilker2022-wt,Yoshimura2023-th} and global budgets on edge-wise quantities \cite{Brown2017-yy}.
More generally, any positive control-dependent rate with a fixed integral can define an alternative notion of time and a rescaled metric.
Among such choices, housekeeping EP is distinguished as the counterpart of the excess EP considered in geometric optimal driving, yielding bounds on the total EP.

Beyond slow driving, finite-time control of the full probability dynamics requires methods such as the Pontryagin principle or dynamic programming~\cite{Muratore-Ginanneschi2013-tc,Zhong2022-ap}; extending the entropic-time approach to thie setting remains an open challenge.

Entropic time has appeared in several other contexts.
In stochastic thermodynamics, random-time transformations have been utilized
\cite{Pigolotti2017-mb,Chun2019-jn}, whereas our entropic time is protocol-determined.
Related entropy-based notions of time have also been investigated in biological aging and growth \cite{Martyushev2025-kp,Tarkhov2024-dt}, as well as in diffusion-model scheduling \cite{Stancevic2025-sc,Foresti2026-sb}.
Our framework of optimal control governed by entropic time thus opens up promising avenues for interdisciplinary physics.

\paragraph{Acknowledgement.}
This work was supported by JSPS KAKENHI (Grant No. 25H01365) and  JST CREST (Grant No. JPMJCR25Q2).
K. S. acknowledges support from JST SPRING (Grant No. JPMJSP2108).



\bibliography{apssamp}

\end{document}